\documentclass[amsmath,amssymb,aps,prd,showkeys,reprint]{revtex4-1}
\usepackage[breaklinks,colorlinks=true]{hyperref}
\usepackage{graphicx,color}

\newcommand\tr{\mathop{\mathrm{Tr}}}
\newcommand\br[1]{\left(#1\right)}
\newcommand\bbr[1]{\left[#1\right]}

\newcommand\veck{\boldsymbol{k}}

\begin{document}

\title{Chiral torsional effect with finite temperature, density and curvature}

\author{Shota~Imaki}
\author{Zebin~Qiu}
\affiliation{Department of Physics, The University of Tokyo, Tokyo 113-0033, Japan}

\begin{abstract}
    We scrutinize the novel chiral transport phenomenon
    driven by spacetime torsion, namely the chiral torsional effect (CTE).
    We calculate the torsion-induced chiral currents
    with finite temperature, density and curvature
    in the most general torsional gravity theory.
    The conclusion complements the previous study on the CTE by including curvature
    and substantiates the relation between the CTE and the Nieh-Yan anomaly.
    We also analyze the response of chiral torsional current
    to an external electromagnetic field.
    The resulting topological current is analogous to that in the axion electrodynamics.
\end{abstract}

\maketitle

\section{Introduction}

A prominent feature of relativistic chiral matter is
the existence of various novel chiral transport phenomena.
Famous examples are the chiral magnetic effect (CME) and
the chiral vortical effect (CVE),
i.e., the generation of electric current along a magnetic field and vorticity respectively
on the condition of chirality imbalance~\cite{Fukushima:2008xe, Son:2009tf, Kharzeev:2015znc, Flachi:2017vlp}.
The chirality imbalance is produced by virtue of the axial anomaly~\cite{Adler:1969gk, *Bell:1969ts}
so that these macroscopic chiral transport phenomena are associated with the underlying quantum anomaly.
It has been demonstrated that the CME is related to the axial anomaly~\cite{Basar:2012gm}.
Meanwhile, the CVE has been regarded as involving also the gravitational anomaly%
~\cite{Delbourgo:1972xb, *Eguchi:1976db, *AlvarezGaume:1983ig, Landsteiner:2011cp, Basar:2013qia}.
As observable manifestations of
the quantum anomaly and topological properties of chiral gauge theories,
chiral transport phenomena have been studied with immense efforts in various physical contexts, e.g.,
quark-gluon plasma in heavy-ion collisions%
~\cite{Kharzeev:2007jp, Burnier:2011bf, Adamczyk:2014mzf, Adamczyk:2015eqo, Khachatryan:2016got, Shi:2017cpu};
topological condensed matter systems such as topological insulators%
~\cite{Qi:2010wf, Ryu:2010ah, Wang:2014dua, Sekine:2015bqa, Higashikawa:2018qsh, Yue:2019hpr}
or Dirac and Weyl semimetals%
~\cite{Zyuzin:2012tv, PhysRevLett.107.186806, Xiong413, Li:2014bha, Chernodub:2015wxa, Burkov:2015hba};
electroweak media in neutron stars%
~\cite{Charbonneau:2009ax, Kaminski:2014jda, Sigl:2015xva, Kaplan:2016drz, Dvornikov:2020qwx},
the primordial universe%
~\cite{Boyarsky:2011uy, Gorbar:2016klv, Pavlovic:2016gac},
or core-collapse supernovae%
~\cite{Yamamoto:2015gzz, Yamamoto:2016xtu, Yamamoto:2020zrs}.

Very recently, a rather new type of chiral transport phenomenon is discovered.
It is induced by the spacetime torsion in the presence of chirality imbalance,
and naturally termed ``chiral torsional effect'' (CTE)~\cite{Khaidukov:2018oat}.
Torsion is a hypothetical spacetime property
in the augmented gravity theory called the Einstein-Cartan gravity,
which has raised great attention among gravity physics
as reviewed by Refs.~\cite{Hehl:1976kj, Shapiro:2001rz}.
It arouses extra research interest that the CTE is supposed
to be connected with the Nieh-Yan anomaly%
~\cite{Nieh:1981xk, *Nieh:1981ww, *Nieh:2007zz}
which depicts the torsional topology of spacetime%
~\cite{Nissinen:2019kld, *Nissinen:2019mkw, *Nissinen:2019wmh, Huang:2019haq, *Huang:2019adx, *Huang:2020ypv}.

Although never observed in real spacetime so far,
torsion can be imitated by a lattice dislocation.
This idea is formulated in lattice field theory and buttressed by numerical computation in Ref.~\cite{Imaki:2019ite}.
In condensed matter,
torsion is realizable in diverse materials like graphene~\cite{deJuan:2010zz},
topological insulators~\cite{Hughes:2011hv, Hughes:2012vg, Parrikar:2014usa} and Weyl semimetal~\cite{Volovik:2013fca, Sumiyoshi:2015eda, You:2016wbd, Huang:2018iys, Ferreiros:2018udw},
where the deformation of the materials acts as torsion effectively.
Especially, Weyl semimetal is an ideal context for the CTE experiments since it bears a chirality imbalance as well.

Despite its profound theoretical significance and promising experimental verifiability,
to the best of our knowledge,
the previous studies of the CTE are incomplete in the sense that
they have neglected curvature effects, or specifically, the spin connection term in the covariant derivative,
and overlooked a certain torsional term allowed in the general torsional gravity Lagrangian.
Besides, the connection between the CTE and the Nieh-Yan anomaly remains not entirely clear.
Firm computation in a complete setup is indispensable
for comprehending the interplay between torsion, curvature and axial anomaly.
Hence in this paper, we decisively calculate the CTE current
at finite temperature, density and curvature
in the most general torsional gravity theory.
In addition, we analyze the current driven by the electromagnetic field in torsional spacetime,
unveiling the impact of torsion on the conventional Maxwell electrodynamics.

This paper is organized as follows.
Sec.~\ref{sec:t} serves as a brief review of torsional gravity.
We introduce the basic notion of torsion
and expound the general form of the coupling between torsion and a fermion.
In Sec.~\ref{sec:j}, we calculate the torsion-induced current.
We first evaluate the current at zero temperature and density
to clarify its relation to the Nieh-Yan's torsional topological invariant,
and then generalize our calculation to finite temperature and density.
In Sec.~\ref{sec:jA},
We analyze the current driven by electromagnetic fields
in the presence of torsion,
and hereby illuminate the analogy between the electrodynamics of the torsional gravity theory to the axion electrodynamics.
The conclusive section \ref{sec:c} presents our summary and outlook.
We adopt the Euclidean spacetime throughout this work.

\section{Torsion}
\label{sec:t}

The standard Einstein gravity theory assumes the symmetry of affine connection
$\Gamma^\lambda_{\;\; \mu \nu} = \Gamma^\lambda_{\;\; \nu \mu}$.
Together with the metricity condition, this assumption leads one to identify the affine connection
with the Christoffel symbol determined solely by metric:
\begin{align}
    \Gamma^\lambda_{\;\; \mu \nu}
    & = \frac{1}{2} \Delta^{\alpha \beta \gamma}_{\rho \mu \nu} \,
        g^{\rho \lambda} \partial_{\alpha} g_{\beta \gamma} \,,
\end{align}
with a permutation symbol
$\Delta^{\alpha \beta \gamma}_{\mu \nu \rho} \equiv
    \delta^\alpha_\rho \delta^\beta_\mu \delta^\gamma_\nu
    + \delta^\alpha_\nu \delta^\beta_\mu \delta^\gamma_\rho
    - \delta^\alpha_\mu \delta^\beta_\nu \delta^\gamma_\rho$.
By contrast, the Einstein-Cartan gravity theory relaxes the assumption of a symmetric affine connection,
allowing for an antisymmetric part termed ``torsion'':
\begin{align}
    T^\lambda_{\;\; \mu \nu}
    & \equiv \tilde \Gamma^\lambda_{\;\; \mu \nu} - \tilde \Gamma^\lambda_{\;\; \nu \mu} \,.
    \label{eq:T}
\end{align}
We henceforth attach tilde in denoting quantities containing torsion.
Affine connection itself is not a tensor,
but torsion is, thus qualified as a physical quantity.
Provided the metricity condition, the relation between
$\tilde \Gamma^\lambda_{\;\; \mu \nu}$ and $\Gamma^\lambda_{\;\; \mu \nu}$ reads
\begin{align}
    \tilde \Gamma_{\;\; \mu \nu}^\lambda
    & = \Gamma_{\;\; \mu \nu}^\lambda -
        \frac{1}{2} \Delta^{\alpha \beta \gamma}_{\rho \mu \nu}
        g^{\lambda \rho} T_{\alpha \beta \gamma} \,.
    \label{eq:GammaT}
\end{align}
Equation~\eqref{eq:GammaT} demonstrates that the spacetime features
two independent intrinsic properties, metric and torsion.

Correspondingly, the covariant derivative of a spinor field
comprises an extra term embodying the coupling of torsion with a fermion:
\begin{align}
    \tilde \nabla_\mu \psi
    & \equiv
        \nabla_\mu\psi
        + \frac{i}{16} \Delta^{\alpha \beta \gamma}_{\nu \rho \mu} \,
        T_{\alpha \beta \gamma} \,
        (e_a^\nu e_b^\rho - e_b^\nu e_a^\rho)
        \sigma^{a b}\,
        \psi \,,
    \label{eq:nablaT}
\end{align}
where $e_m^\mu$ is the vierbein satisfying the orthonormal relations
$e_m^\mu e_{\mu n} = \delta_{m n},\, e_{\mu m} e_\nu^m = g_{\mu \nu}$.
The first term in Eq.~\eqref{eq:nablaT} is
the torsion-free covariant derivative in the Einstein gravity theory:
\begin{align}
    \nabla_\mu \psi
    & \equiv
        \partial_\mu \psi
        + \frac i 2 \omega_{\mu a b} \sigma^{a b} \psi \,,
\end{align}
with $\sigma^{a b} \equiv \frac i 2 [\gamma^a, \gamma^b]$
and the spin connection
\begin{align}
    \omega_{\mu a b}
    & = \frac 1 4
        (e_{b \sigma} \partial_\mu e_a^\sigma - e_{a \sigma} \partial_\mu e_b^\sigma)
        + \frac 1 4
        \Gamma^\alpha_{\;\; \beta \mu}
        (e_a^\beta e_{b \alpha} - e_b^\beta e_{a \alpha}) \,.
\end{align}

With the covariant derivative defined by Eq.~\eqref{eq:nablaT},
we readily write down the Dirac Lagrangian in torsional curved spacetime,
\begin{align}
    \mathcal L_\text{min}
    & = \frac{1}{2} \bar \psi (
            \gamma^\mu \tilde \nabla_\mu - m
        ) \psi
        + \text{h.c.} \,,
    \label{eq:Lmin0}
\end{align}
which is sometimes called the minimal theory.
After some algebra~\cite{Shapiro:2001rz}, we rewrite Eq.~\eqref{eq:Lmin0}
to sort out the torsional contribution:
\begin{align}
    \mathcal L_\text{min}
    & = \bar \psi \bbr{
            \gamma^\mu \Big( \nabla_\mu - \frac 1 8 \gamma_5 S_\mu \Big) - m
        } \psi \,,
    \label{eq:Lmin}
\end{align}
where $S^\mu$ is what we call ``screw torsion'':
\begin{align}
    S_\mu
    & \equiv
        \varepsilon^{\mu \nu \rho \sigma} T_{\rho \nu \sigma}
\end{align}
with $\varepsilon^{\mu \nu \rho \sigma}$ denoting the covariant Levi-Civita tensor.
The most general Lagrangian obeying
covariance, locality, renormalizability and parity symmetry
allows for another type of torsional term that we call ``edge torsion'',
\begin{align}
    E_\mu
    & \equiv T^{\alpha}_{\;\; \mu \alpha} \,,
\end{align}
and takes the form of:
\begin{align}
    \mathcal L
    & = \bar \psi [
            \gamma^\mu (\nabla_\mu - \eta_1 \gamma_5 S_\mu - \eta_2 E_\mu
            ) - m
        ] \psi \,.
        \label{eq:L}
\end{align}
The parameters $\eta_1$ and $\eta_2$ are arbitrary real numbers for the general theory,
while the specific choice $\eta_1 = 1/8,\, \eta_2 = 0$
recovers the minimal theory \eqref{eq:Lmin}.

Two features of the Lagrangian \eqref{eq:L} play essential roles in later computation.
Firstly, the torsional terms are entirely separated.
Thus we can conveniently define the perturbation
away from the torsion-free theory that
corresponds to the choice $\eta_1 = \eta_2 = 0$.
Secondly, the edge torsion couples to a fermion
in the same way as a $\mathrm{U}(1)$ gauge field.
It enables us to easily encompass an external electromagnetic field
by combining it with the edge torsion:
\begin{align}
    A^\prime_\mu
    & \equiv A_\mu + \eta_2 E_\mu \,.
    \label{eq:Aprime}
\end{align}
In this way, we consider $E_\mu$ together with the electromagnetic field in Sec.~\ref{sec:jA}.
Until then we turn off $A^\prime_\mu$ for simplicity.

\section{Torsion-induced Current}
\label{sec:j}

We aim to evaluate the torsion-induced chiral current
in the most general theory \eqref{eq:L}
with metric and torsion treated as background fields.
Our calculation starts from the following vacuum or thermal expectation value of the chiral current:
\begin{align}
    J_\pm^\mu
    & = \langle \bar\psi \gamma^\mu P_\pm \psi \rangle \,,
    \label{eq:jdef}
\end{align}
where ``$+$'' and ``$-$'' stand for right-handedness and left-handedness respectively,
and $P_\pm \equiv \frac 1 2 (1 \pm \gamma_5)$ denotes the chiral projector.

Throughout the present section, as explained above,
the electromagnetic field together with the edge torsion, $A^\prime_\mu$,
is shut down, and the screw torsion $S_\mu$ is
disposed as a perturbation to the linear order.
In parallel, the effect of curvature is also kept to the leading order
in terms of the curvature tensor $R_{\mu \nu \rho \sigma}$.

The chiral current $J_\pm^\mu$ is calculated in two different setups.
The result at zero temperature and density is achieved in Sec.~\ref{sec:jzero}.
The axial current, in this case, depends on the ultraviolet cutoff
and its divergence proves related to the Nieh-Yan topological invariant.
Then the generalization to finite temperature and density is accomplished in Sec.~\ref{sec:jfinite}.
The chiral current relies on the interplay between torsion and curvature,
and exhibits a distinctive dependence on temperature and density
in contrast to the CME and the CVE.

\subsection{Zero temperature and density}
\label{sec:jzero}

At zero temperature and density,
given that the screw torsion $S_\mu$ is an axial vector,
the vector current vanishes at $\mathcal{O}(S_\mu)$.
We therefore focus on the axial current
\begin{align}
    J_5^\mu
    & = \langle \bar \psi \gamma^\mu \gamma_5 \psi \rangle \,.
\end{align}
We calculate it as the trace involving the propagator,
in a similar way to Ref.~\cite{Imaki:2019ite}.
The perturbative expansion with respect to the screw torsion
gives rise to
\begin{align}
    J_5^\mu
    & = - \eta_1 \tr (\gamma^\mu \gamma_5 G \gamma^\nu \gamma_5 G) S_\nu
        + \mathcal O(S_\mu^2,\, \partial^2 S_\mu) \,,
    \label{eq:jper}
\end{align}
with $G$ representing the torsion-free propagator and
$\tr$ standing for the trace over both Dirac indices and coordinate space.
We make two remarks about our power counting.
Firstly, given the symmetry property of the curvature tensor,
the torsion-independent part vanishes at zero temperature and density,
as also pointed out in Ref.~\cite{Flachi:2017vlp}.
Secondly, from the perspective of parity,
one can understand that the first-order derivative of $S^\mu$
does not contribute to the axial current.

To simplify our computation,
we employ the Riemann normal coordinate
around the point $x$ at which the current is evaluated.
In this coordinate system,
the Christoffel symbol $\Gamma^\mu_{\;\; \nu \rho}$ vanishes at $x$
and the $\gamma$-matrices are those in flat spacetime.
After the transformation into momentum $k$-space,
the propagator at the coincidental point acquires the following form
according to Ref.~\cite{Bunch:1979uk, *Parker:2009uva}:
\begin{align}
    G(x, x^\prime \to x)
    & = \int \frac{d^4 k}{(2 \pi)^4} \;
        (i \gamma^\mu k_\mu + m) \,
        \mathcal G(k) \,,
        \label{eq:Gzero}
\end{align}
where the function $\mathcal G(k)$ includes curvature effects in a perturbative way:
\begin{align}
    \mathcal G(k)
    & = - \Bigg [
            1 - \br{
                A_1
                + i A_{1 \alpha} \frac{\partial}{\partial k_\alpha}
                - A_{1 \alpha \beta} \frac{\partial^2}{\partial k_\alpha \partial k_\beta}
            }
            \frac{\partial}{\partial m^2} \notag \\
            & \quad \quad \quad
            + A_2 \br{\frac{\partial}{\partial m^2}}^2
        \Bigg ]
        \frac{1}{k^2 + m^2}
        + \cdots \,.
        \label{eq:gzero}
\end{align}
The first coefficient $A_1$ is proportional to the scalar curvature,
\begin{align}
    A_1 = \frac{R}{12} \,.
    \label{eq:A1}
\end{align}
The subsequent coefficients, $A_{1 \alpha}$, $A_{1 \alpha \beta}$, $A_2$ and so forth,
consist of higher orders of curvature or derivatives thereof.
One can refer to Ref.~\cite{Bunch:1979uk, *Parker:2009uva} for the specific value of them
but we focus on the leading-order curvature effect
so that $A_1$ suffices.

Inserting Eqs.~\eqref{eq:Gzero} and \eqref{eq:gzero}
into Eq.~\eqref{eq:jper} and taking the trace over Dirac indices yield:
\begin{align}
    J_5^\mu
    & = 2 \eta_1 S^\mu
        \int^\Lambda \frac{d^4 k}{(2 \pi)^4} \;
        (2 m^2 - k^2) \, \mathcal G^2(k) \,.
    \label{eq:sumintegral}
\end{align}
We have introduced the ultraviolet cutoff $\Lambda$
so as to figure out the dependence of the axial current on $\Lambda$,
which is also implied in
Refs.~\cite{Nissinen:2019kld, *Nissinen:2019mkw, *Nissinen:2019wmh, Huang:2019haq, *Huang:2019adx, *Huang:2020ypv}.
With detailed computation left in Appendix~\ref{sec:Izero},
we present the conclusive result as:
\begin{align}
    J_5^\mu
    &= \frac{\eta_1}{8 \pi^2} S^\mu \notag \\
        &
        \cdot \bbr{
            - \Lambda^2 - 3 m^2 + \frac{5}{12} R
            + \Big( 4 m^2 - \frac R 6 \Big)
            \log \Big(1 + \frac{\Lambda^2}{m^2} \Big)
        } \,.
    \label{eq:j5mu}
\end{align}

Let us examine the axial anomaly indicated by Eq.~\eqref{eq:j5mu}.
To this end, we take the massless limit $m \to 0$.
Furthermore, since the curvature is independent of torsion
and irrelevant to our interest here,
we rightfully take $R_{\mu \nu \rho \sigma} = 0$.
Then the axial current reads
\begin{align}
    J_5^\mu
    & = - \frac{\eta_1 \Lambda^2}{8 \pi^2} S^\mu \,.
    \label{eq:j5cut}
\end{align}
Accordingly, the divergence of the axial current takes the form of~\cite{Nieh:2007zz}
\begin{align}
    \partial_\mu J_5^\mu
    & = \frac{\eta_1 \Lambda^2}{8 \pi^2} \,
        \varepsilon^{\mu \nu \rho \sigma} \,
        T_{\;\; \mu \nu}^\alpha T_{\alpha \rho \sigma} \,.
    \label{eq:divj5}
\end{align}
In fact, the volume integral of the divergence
is proportional to Nieh-Yan's topological invariant~\cite{Nieh:1981ww},
\begin{align}
    N_\text{NY}
    & = \int d^4 x \;
        \varepsilon^{\mu \nu \rho \sigma} \,
        T^\alpha_{\;\; \mu \nu} T_{\alpha \rho \sigma} \,,
    \label{eq:NY}
\end{align}
which characterizes the torsional topology of spacetime.
The relation \eqref{eq:divj5} is referred to as the Nieh-Yan anomaly%
~\cite{Nissinen:2019kld, *Nissinen:2019mkw, *Nissinen:2019wmh, Huang:2019haq, *Huang:2019adx, *Huang:2020ypv}
in that the right-hand side has an anomalous nature 
and the left-hand side embodies Nieh-Yan's topological invariant.

It is noteworthy that in a general sense,
the divergence of the axial current in a torsional curved spacetime
receives other contributions in addition to Eq.~\eqref{eq:divj5},
which we are nevertheless unable to capture under our truncation scheme.
For instance, Nieh-Yan's topological invariant
should own the Pontryagin form of the curvature~\cite{Nieh:1981ww}
in accompany with Eq.~\eqref{eq:NY},
which is at the second order of the curvature tensor.
Besides, as firstly indicated in Ref.~\cite{Obukhov:1982da, *Obukhov:1983mm},
there is $\Lambda$-independent torsional contribution
to the axial anomaly from higher orders of torsion and its derivative.
To grasp this, one shall extend our analysis
to include higher-order terms of curvature and torsion.

\subsection{Finite temperature and density}
\label{sec:jfinite}

We now generalize to
the chiral current \eqref{eq:jdef} at finite temperature $T$,
vector chemical potential $\mu$ and axial chemical potential $\mu_5$.
For such purpose, we resort to the Matsubara formalism.
We impose the stationary condition of metric, i.e.,
all metric components are time-independent and
the temporal components are space-independent,
which justifies the standard Matsubara formalism.
For simplicity, we consider a massless fermion with $m = 0$.

We observe from the Lagrangian \eqref{eq:L} that
the temporal component of the screw torsion couples to a fermion
in an identical way with the axial chemical potential.
Thus we absorb it into a redefined axial chemical potential:
\begin{align}
    \mu_5^\prime
    & \equiv \mu_5 + \eta_1 S_\tau \,.
\end{align}
Then without loss of generality,
we specify the screw torsion to be pure space-like, and
further direct it along the $\mathrm z$-axis as $S_\mu = S_z \hat z$
on account of spherical symmetry.
One can manifest that only the $\tau$- and $\mathrm z$-components
of the current \eqref{eq:jdef} are nonvanishing.
Since the $\tau$-component does not depend on $S_z$ at the linear order,
we focus on the $\mathrm z$-component,
\begin{align}
    J_\pm^z
    & = \langle \bar\psi \gamma^z P_\pm \psi \rangle \,.
    \label{eq:jpmfinite}
\end{align}

It is straightforward to prove that the current \eqref{eq:jpmfinite}
can be evaluated by a similar formula to Eq.~\eqref{eq:jper}
with the momentum $k_\mu$ therein replaced by
\begin{align}
    K_{\pm \mu}
    & \equiv (\veck,\, \omega_n + i \mu_\pm) \,,
\end{align}
with the Matsubara frequencies $\omega_n \equiv 2 \pi T (n + \frac 1 2)$
and the chiral chemical potential $\mu_\pm \equiv \mu \pm \mu_5'$.
To the linear order, the chiral current is expressed as
\begin{align}
    J_\pm^z
    & = - \eta_1 \tr (\gamma^z P_\pm G_\pm \gamma^z \gamma_5 G_\pm) S^z
        + \cdots \,,
        \label{eq:jpmdef}
\end{align}
where $G_\pm$ is given by
\begin{align}
    G_\pm
    & = T \sum_{\omega_n} \int \frac{d^3 \veck}{(2 \pi)^3} \;
        i \gamma^\mu K_{\pm \mu} \,
        \mathcal G_\pm(k)
        \label{eq:G}
\end{align}
with the perturbative expansion of $\mathcal G_\pm (k)$
being formally similar to Eq.~\eqref{eq:gzero},
\begin{align}
    \mathcal G_\pm(k)
    & = - \frac{1}{K_\pm^2}
        - \frac{R}{12} \frac{1}{(K_\pm^2)^2}
        + \cdots \,.
        \label{eq:g}
\end{align}

Applying the formulas \eqref{eq:G} and \eqref{eq:g}
to the expression \eqref{eq:jpmdef} and
carrying out the Dirac trace,
we boil the computation down to the following sum-integral:
\begin{align}
    J_\pm^z
    & = \pm 2 \eta_1 S^z \,
        T \sum_{\omega_n} \int \frac{d^3 \veck}{(2 \pi)^3} \;
        (2 k_z^2 - K_\pm^2) \, \mathcal G_\pm^2(k) \,.
        \label{eq:sumintegralfinite}
\end{align}
Now that we are interested in the dependence of $J_\pm^z$
on temperature and density rather than the ultraviolet scale,
we calculate the integral with dimensional regularization and subtract the divergence according to the modified minimal subtraction scheme.
After the computation of the sum-integral detailed in Appendix~\ref{sec:Ifinite},
we obtain the final result:
\begin{align}
    J_\pm^z
    & = \pm \frac{\eta_1 R}{96 \pi^2} \, S^z \,
        F \Big( \frac{\mu_\pm}{2 \pi T} \Big) \,.
    \label{eq:jpm}
\end{align}
The dependence on temperature and density
is expressed utilizing digamma function $\psi(z)$ as
\begin{align}
    F(z)
    & = \psi \Big( \frac12 + i z \Big)
        + \psi \Big( \frac12 - i z \Big),
    \label{eq:F}
\end{align}
which is depicted in Fig.~\ref{fig:f}.

\begin{figure}[t]
    \includegraphics[width=1.0\linewidth]{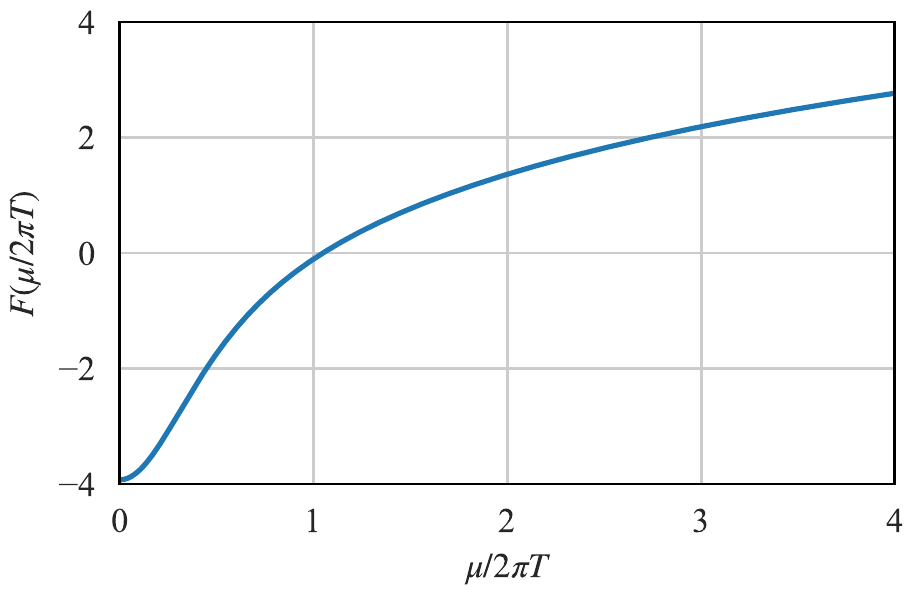}
    \caption{
        \label{fig:f}
        The dependence of the function $F$ in
        Eq.~\eqref{eq:jpm} on temperature and density.
    }
    \label{fig:f}
\end{figure}

We remark that the result \eqref{eq:jpm} should not be directly compared
with that in zero temperature and density \eqref{eq:j5mu},
because the result of $J_\pm^\mu$ would be changed
by altering the order of taking the three limits,
$T \to 0$, $\mu_\pm \to 0$, and $m \to 0$.
For example in Eq.~\eqref{eq:jpm}, the $\mu_\pm \to 0$ limit
can be taken straight while the $T \to 0$ limit should be analyzed
through the asymptotic expansion, and apparently the results bear
different coefficients of $RS^z$.
Discussion on the $T \to 0$ limit is provided in Appendix~\ref{sec:Ifinite}.
Moreover, during the dimensional regularization,
an infinite portion in $J_\pm^z $ exists
as the counterpart of the $\Lambda$-dependent term in Eq.~\eqref{eq:j5mu},
but has already been subtracted, thus absent in Eq.~\eqref{eq:jpm}.

\section{Torsional Electrodynamics}
\label{sec:jA}

Now we come to the study of the current response
to an external electromagnetic field in the presence of torsion.
It is worth reminding that we combine the edge torsion with
the electromagnetic field as
\begin{align}
    A^\prime_\mu
    & \equiv A_\mu + \eta_2 E_\mu \,.
\end{align}
Hence our analysis in this section accounts for
the current driven by the edge torsion as well.
For simplicity, we confine our study to
the massless fermion on a flat metric with zero chemical potential.
We also assume the screw torsion to be stationary and homogeneous.
Under these assumptions,
we can perform an axial transformation
\begin{align}
    \psi(x)
    \to \exp(- i \eta_1 \gamma_5 S_\mu x^\mu) \psi(x)
\end{align}
to eliminate $S_\mu$ from the fermionic sector of the Lagrangian.
This transformation meanwhile yields the following anomalous term
in the gauge sector:
\begin{align}
    S_{\text{anom}}
    & = \frac{\eta_1}{4 \pi^2} \int d^4x \;
        A'_{\mu} S_\nu \tilde F^{\mu \nu} \,,
    \label{eq:Sanom}
\end{align}
where $F_{\mu \nu} \equiv \partial_\mu A^\prime_\nu - \partial_\nu A^\prime_\mu$
and $\tilde F^{\mu \nu} \equiv \frac 1 2 \varepsilon^{\mu \nu \rho \sigma} F_{\rho \sigma}$.
Remarkably, Eq.~\eqref{eq:Sanom} is formally the same as the action of axion electrodynamics
and the screw torsion plays the role of the derivative of the vacuum angle:
$S_\mu \sim \partial_\mu \theta$.

The functional derivative of the action \eqref{eq:Sanom} with respect to $A_\mu$
gives rise to the vector current,
\begin{align}
    J^\mu
    & = \frac{\eta_1}{2 \pi^2}
        S_\nu \tilde F^{\mu \nu} \,.
    \label{eq:jem}
\end{align}
This equation summarizes multiple torsion-induced phenomena.
The temporal component represents an anomalous charge density
\begin{align}
    n
    & = \frac{\eta_1}{2 \pi^2}
        \boldsymbol S \cdot \boldsymbol B \,,
        \label{eq:witten}
\end{align}
resembling the Witten effect~\cite{Witten:1979ey},
in which magnetic flux traversing the gradient of the vacuum angle induces the extra charge.
Thus we entitle Eq.~\eqref{eq:witten} the torsional Witten effect.
On the other hand, the spatial component of the current reads:
\begin{align}
    \boldsymbol J
    & = \frac{\eta_1}{2 \pi^2}
        (S_\tau \boldsymbol B + \boldsymbol S \times \boldsymbol E) \,.
\end{align}
The first term is the torsional realization of the chiral magnetic effect~\cite{Fukushima:2008xe}
in which $S_\tau$ acts as the axial chemical potential.
We thereupon designate it as the torsional magnetic effect.
The second term is a current perpendicular to the electric field,
which we name the torsional Hall effect after the anomalous Hall effect~\cite{Jungwirth:2002zz, *Fang:2003ir, Haldane:2004zz}.

As the parity dual of the vector current \eqref{eq:jem},
the axial current is derived in parallel from the anomalous action \eqref{eq:Sanom} as
\begin{align}
    J_5^\mu
    & = \frac{\eta_2}{4 \pi^2}
        E_\nu \tilde F^{\mu \nu} \,.
\end{align}
Given that the torsion is mimicked by lattice dislocation~\cite{Imaki:2019ite},
this relation would be suggestive for condensed matter experiments
about creating chirality imbalance without axial chemical potential.

\section{Conclusion}
\label{sec:c}

We calculate the torsion-induced current at finite temperature, density and curvature for the general Einstein-Cartan gravity theory.
The axial current at zero temperature and density
reveals the relation between the CTE and Nieh-Yan's topological invariant.
The chiral current at finite temperature and density
features a rather nontrivial dependence on temperature and density,
distinguished from the quadratic dependence on $T$ and $\mu_\pm$ in the CVE.

Our work has not only theoretical significance but also
phenomenological implications.
It has been proposed that torsion can be realized as lattice dislocation,
indicating that the torsion-induced current is experimentally verifiable.
The interaction between torsion and electromagnetic field
demonstrates torsion as an alternative to the axial chemical potential
for the production of chirality imbalance,
heralding broader physical contexts for the study of chiral transport phenomena.
The analogy between torsional electrodynamics and axion electrodynamics
substantiates that novel topological effects in the latter
can exist in a torsional spacetime even without a vacuum angle.
One interesting example is the recently discovered axionic Casimir force
that proves anomalously repulsive in Ref.~\cite{Fukushima:2019sjn}.

Based on this paper, several future directions await us to explore.
For example, we have truncated the result to the leading order
of both torsion and curvature.
The generalization to higher orders would fully
clarify the relation between the torsion-induced current and
the axial anomaly in the Einstein-Cartan gravity theory.
Also, we have treated the torsion as a background field and
the extension to dynamical torsion would be a challenging yet intriguing future task.

\begin{acknowledgments}
The authors thank Kenji~Fukushima, Kazuya~Mameda and Arata~Yamamoto for beneficial discussions.
S.~I.~was supported by Grant-in-Aid for JSPS Fellows Grant Number 19J22323.
\end{acknowledgments}

\appendix

\section{Integrals in zero temperature and density}
\label{sec:Izero}

We supply details for the calculation of Eq.~\eqref{eq:j5mu}.
Up to the leading orders of mass and curvature,
Eq.~\eqref{eq:sumintegral} involves the following integrals
calculated with hard the cutoff at $k^2 = \Lambda^2$:
\begin{align}
    & \int^\Lambda \frac{d^4 k}{(2 \pi)^4} \frac{1}{(k^2 + m^2)^3}
        = \frac{1}{32 \pi^2 m^2} \,,
    \\
    & \int^\Lambda \frac{d^4 k}{(2 \pi)^4} \frac{1}{(k^2 + m^2)^2}
        = \frac{1}{16 \pi^2} \bbr{
                \log \Big(1 + \frac{\Lambda^2}{m^2} \Big) - 1
            } \,,
    \\
    & \int^\Lambda \frac{d^4 k}{(2 \pi)^4} \frac{k^2}{(k^2 + m^2)^3}
        = \frac{1}{16 \pi^2} \bbr{
                \log \Big(1 + \frac{\Lambda^2}{m^2} \Big) - \frac 3 2
            } \,,
    \\
    & \int^\Lambda \frac{d^4 k}{(2 \pi)^4} \frac{k^2}{(k^2 + m^2)^2} \notag \\
        & \quad
        = \frac{1}{16 \pi^2} \bbr{
                \Lambda^2 - 2 m^2 \log \Big(1 + \frac{\Lambda^2}{m^2} \Big) + m^2
            } \,.
\end{align}
These formulas lead us to the result \eqref{eq:j5mu}.

\section{Integrals in finite temperature and density}
\label{sec:Ifinite}

We provide a concrete derivation of Eq.~\eqref{eq:jpm}.
With the expansion \eqref{eq:g} applied,
the current \eqref{eq:sumintegralfinite} equals
\begin{align}
    J_\pm^z
    & = \pm 2 \eta_1 S^z
           \bbr{
               2 I_2^z - I_1 + \frac R 6 (2 I_3^z - I_2)
           } \,,
    \label{eq:appJpm}
\end{align}
where we have defined for convenience the following sum-integrals:
\begin{align}
    I_n
    \equiv
        T \sum_{\omega_n} \int \frac{d^3 \veck}{(2 \pi)^3} \;
        \frac{1}{(K_\pm^2)^n} \,,
    \\
    I_n^z
    \equiv
        T \sum_{\omega_n} \int \frac{d^3 \veck}{(2 \pi)^3} \;
        \frac{k_z^2}{(K_\pm^2)^n} \,.
\end{align}
We adopt the dimensional regularization
$d^3 \veck / (2 \pi)^3 \to M^{3 - d} d^d \veck / (2 \pi)^d$
with the number of dimensions $d = 3 - 2 \epsilon$ and the scale parameter $M$.
Then we carry out the momentum integrals:
\begin{align}
    I_n
    & = \frac{M^{3 - d} \Gamma(n - \frac d 2)}{(4 \pi)^{\frac d 2} \Gamma(n)} \,
        T \sum_{\omega_n} (K_{\pm \tau}^2)^{\frac d 2 - n} \,,
    \\
    I_n^z
    & = \frac{1}{2 (n - 1)} I_{n - 1} \,.
\end{align}
After some algebra, the Matsubara sum amounts to
\begin{align}
    & \sum_{\omega_n} (K_{\pm \tau}^2)^{\frac d 2 - n} \notag \\
    & = (2 \pi T)^{d - 2 n}
        \bbr{
            \zeta \Big( -d + 2 n,\, \frac 1 2 + i \frac{\mu_\pm}{2 \pi T} \Big)
            + \text{c.c.}
        },
    \label{eq:In}
\end{align}
where $\zeta(z,\, a)$ denotes the Hurwitz zeta function.
The integrals $I_1$ and $I_2^z$ have no divergence at $\epsilon = 0$
and read directly:
\begin{align}
    I_1
    & = - \frac{\mu_\pm^2}{8 \pi^2}
        - \frac{T^2}{24}
        + O(\epsilon) \,,
    \label{eq:resultI1}
    \\
    I_2^z
    & = - \frac{\mu_\pm^2}{16 \pi^2}
        - \frac{T^2}{48}
        + O(\epsilon) \,.
    \label{eq:resultI2z}
\end{align}
On the other hand, the integrals $I_2$ and $I_3^z$
diverge at $\epsilon = 0$ and therefore request regularization.
We exploit the Laurent series expansion of the Hurwitz zeta function:
\begin{align}
    \zeta(1 + 2 \epsilon,\, z)
    & = \frac{1}{2 \epsilon}
        - \psi(z)
        + \mathcal O (\epsilon) \,.
    \label{eq:hurwitzdigamma}
\end{align}
In this way, we derive
\begin{align}
    I_2
    & = \frac{1}{16 \pi^2}
        \bbr{
            \frac{1}{\bar \epsilon}
            + 2 \log \Big( \frac{M}{4 \pi T} \Big)
            - F \Big( \frac{\mu_\pm}{2 \pi T} \Big)
        }
        + O(\epsilon) \,,
    \label{eq:Iep}
\end{align}
where the definition of the function $F(z)$ has been clarified in Eq.~\eqref{eq:F}
and the constant is defined as
\begin{align}
    \frac{1}{\bar\epsilon}
    & = \frac{1}{\epsilon}
        - \gamma_{\text E}
        + \log (4 \pi) \,.
\end{align}
Following the modified minimal subtraction scheme,
we subtract the infinity as well as the logarithmic term in Eq.~\eqref{eq:Iep},
and obtain the final result:
\begin{align}
    I_2
    & = - \frac{1}{16 \pi^2} F \Big( \frac{\mu_\pm}{2 \pi T} \Big) \,,
    \label{eq:resultI2}
    \\
    I_3^z
    & = - \frac{1}{64 \pi^2} F \Big( \frac{\mu_\pm}{2 \pi T} \Big) \,.
    \label{eq:resultI3z}
\end{align}
Eventually, one can easily attain the chiral current \eqref{eq:jpm}
by plugging the sum-integrals
\eqref{eq:resultI1}, \eqref{eq:resultI2z}, \eqref{eq:resultI2} and \eqref{eq:resultI3z}
into the formula \eqref{eq:appJpm}.

Notably, though temperature $T$ appears in the denominator of the variable of $F(z)$,
taking the zero-temperature limit $T \to 0$ does not incur singularity,
because digamma function converges for a variable with the large imaginary part.
To elaborate this point, we perform the asymptotic expansion of digamma function
\begin{align}
\psi(z)=\log z-\frac{1}{2z}-\frac{1}{12z^{2}}+\cdots \,,
\end{align}
which leads to
\begin{align}
I_{2} = \frac{1}{16\pi^{2}}
        \left(\frac{1}{\bar\epsilon}
        +2\log\frac{\pi M}{\mu_{\pm}}
        +\frac{\pi^{2}T^{2}}{3\mu^{2}_{\pm}}
        \right)
        +\cdots \,.
\end{align}
This equation illuminates the proper way to examine the low or zero-temperature limit of
our result \eqref{eq:jpm}. By comparison, one can analyze the small density limit via
the Taylor expansion of Eq.~\eqref{eq:jpm} with respect to $\mu_{\pm}$ straightforwardly.

\bibliography{torsion}
\bibliographystyle{apsrev4-1}

\end{document}